\documentclass[preprint,aps,prb,showpacs]{revtex4}

\usepackage{graphicx}
\usepackage{bm}

\begin{document}

\title{Bethe-Salpeter Equation Calculations of Core Excitation Spectra}

\author{J. Vinson}
\affiliation{Dept.\ of Physics, Univ.\ of Washington, Seattle, WA 98195}

\author{E. L. Shirley }
\affiliation{National Institute of Standards and Technology (NIST), Gaithersburg,
MD 20899}

\author{J. J. Rehr}
\affiliation{Dept.\ of Physics, Univ.\ of Washington, Seattle, WA 98195}

\author{J. J. Kas}
\affiliation{Dept.\ of Physics, Univ.\ of Washington, Seattle, WA 98195}

\date{\today}

\begin{abstract}
We present a hybrid approach for GW/Bethe-Salpeter Equation (BSE) calculations
of core excitation spectra, including x-ray absorption (XAS), electron energy
loss spectra (EELS), and non-resonant inelastic x-ray scattering (NRIXS).
The method is based on {\it ab initio}
wavefunctions from the plane-wave pseudopotential code ABINIT;
atomic core-level states and projector augmented wave (PAW) transition
matrix elements; the NIST core-level BSE solver;
and a many-pole GW self-energy model to account for
final-state broadening and self-energy shifts. Multiplet effects
are also accounted for.  The approach is implemented using an
interface dubbed OCEAN (Obtaining Core Excitations using ABINIT and NBSE).
To demonstrate the utility of the code we present results for the K-edges 
in LiF as probed by XAS and NRIXS, the K-edges of KCl as probed by XAS, 
the Ti L$_{2,3}$-edge in SrTiO$_3$ as probed by XAS, and the Mg L$_{2,3}$-edge 
in MgO as probed by XAS. We compare the results to experiments and 
results obtained using other theoretical approaches.

\end{abstract}
\pacs{78.70.Dm, 78.20.Bh, 71.15.Qe}

\maketitle

\section{Introduction}
 Recently there has been considerable progress in the theory of
optical response beyond the independent-particle approximation.\cite{Onida02} 
For example, methods based on time-dependent density-functional theory
(TDDFT) and the GW/Bethe-Salpeter Equation (GW/BSE) approach have been extensively
studied.\cite{Onida02,ChelikowskyXX,BertschYabana, Octopus}
While computationally simpler than the BSE, TDDFT is currently
limited by approximations to the exchange-correlation functional.
On the other hand, the GW/BSE approach includes an explicit treatment
of quasi-particle effects within Hedin's GW self-energy
approximation\cite{PhysRev.139.A796} and particle-hole interactions,
both of which are often crucial to a quantitative treatment. In the GW
approximation the electron self-energy is related to the product of the one-electron
Green's function and screened Coulomb interaction, which are respectively 
denoted by symbols $G$ and $W$.
A number of codes based on these approaches have been developed both
for periodic materials\cite{lawler:205108,RevModPhys.73.515,PhysRevLett.75.818} 
and other systems.\cite {Prange09,Chelikowsky09}

 Calculations of core-level spectra, on the other hand, pose additional
theoretical challenges. Core-hole effects, energy-dependent
damping, self-energy shifts, and atomic multiplet effects all complicate
the theory.  Consequently relatively few GW/BSE treatments presently
exist.\cite{Shirley200477,0953-8984-21-10-104205,ComptesRendus09}
To address these challenges, we present here a hybrid GW/BSE
approach for periodic systems encompassing x-ray absorption spectra
(XAS) and related core-excitation spectra. 
Our BSE Hamiltonian also accounts for atomic-multiplet effects
in the spectra. Since our implementation includes self-consistent
potentials for a given system, it improves on multiplet approximations
that rely on crystal-field parameters. Also, although our
approach is designed for
periodic systems, aperiodic systems can be modeled using supercells.
However, the method is limited to a range of order 10$^2$ eV above a given core threshold.
Thus the method is complementary to the real-space Green's function (RSGF)
approach  for core spectra,\cite{Ankudinov05,Rehr05,ComptesRendus09} 
which is applicable over a very broad spectrum of excitation energies
up to about 10$^4$ eV.  Though formally equivalent to the GW/BSE of 
this work, the RSGF implementation uses finite clusters and spherical
scattering potentials which can be inaccurate near threshold.

GW/BSE calculations of core-level spectra rely on a number of 
theoretical and many-body considerations.  A key ingredient is 
the screened electron-core hole interaction.  In the BSE, this interaction is
typically treated via linear response, which in itself can be a demanding task. 
In contrast, many current calculations of core-excitation spectra are based
on effective independent-electron models with various {\it ad hoc}
treatments of the screening effects.\cite{PhysRevB.58.7565,StoBE,Soininen2001} 
For example, many implementations use a final-state Hamiltonian with a
self-consistently screened core hole,
\cite{PhysRevB.58.7565,Schwarz2003259,RevModPhys.73.515,EBERT_SPRKKR,paratec02,Soininen2001}
as in the final-state rule. 
A second important ingredient is a complex, energy-dependent 
self-energy to account for final-state self-energy shifts
and damping.  As shown below, such quasi-particle effects are important 
for a quantitative account of peak positions, heights, and widths in the
spectra. Although small near an absorption edge and often neglected,
final-state damping becomes particularly important above about 10 eV
where interband and plasmon excitations can be important.
A third important ingredient is an account of intra-atomic interactions that
lead to multiplet effects in the spectra.  Finally, the approach should include 
accurate self-consistent potentials and a basis that encompasses the
near edge structure over a range of about 10 eV to 100 eV.

In order to address the above considerations
our core-level GW/BSE approach is based on five key elements:
1) Orbitals for occupied and unoccupied Kohn-Sham
levels from a self-consistent plane-wave pseudopotential code;
2) atomic core-level states and projector augmented wave (PAW) transition
matrix elements;
3) the NIST core-level Bethe-Salpeter Equation solver (NBSE);
4) a many-pole GW self-energy model (MPSE) to approximate final-state 
broadening and self-energy shifts; and finally
5) atomic-multiplet effects are included through the
inclusion of core hole spin-orbit splitting and atomic multipole interactions.  
 In the present implementation ABINIT\cite{Gonze2002478} is used
for  the Kohn-Sham wave functions; however, this is not a strict
restriction and the code can be adapted to other plane-wave pseudopotential
codes. For the multiplet calculations, the electron-core hole wave function
is expressed in terms of electron PAW function-core hole product states,
with the Coulomb interaction matrix elements between these product states
calculated using an atomic structure program.   Vibrational damping
is neglected as such such damping effects are generally small
in the near-edge regime.

Our hybrid GW/BSE approach provides a first-principles method
for calculations of near-edge spectra, including x-ray absorption spectra
(XAS), electron energy-loss spectra (EELS), and non-resonant inelastic
scattering spectra (NRIXS), also known as x-ray Raman spectra, at finite
momentum-transfer ${\bf q}$. The core-BSE calculations are quite efficient
compared to those for optical spectra based, e.g., on AI2NBSE,
since the core-level subspace is smaller than the valence band manifold. 
The approach is implemented
using an interface dubbed OCEAN (Obtaining Core Excitation using ABINIT and
NBSE), which generates relevant input files for the various modules and
serves as a driver for all steps of the calculation.
This interface is a generalization to core excitations of AI2NBSE, a recently
developed interface for valence excitations.\cite{lawler:205108}
In the current
implementation, the OCEAN package can handle up to about 50 atoms per unit
cell and spectra up to about 100 eV above threshold. 

The remainder of this paper is as follows.  Section\ II. \ summarizes the theory
underlying our approach. Section\ III.\ presents illustrative
results for the K-edges in LiF and KCl,
the Mg L$_{2,3}$-edge in MgO, and the Ti L$_{2,3}$-edge in SrTiO$_{3}$. 
Finally, Section\ IV.\ presents a summary and prospects for future
development.

\section{Theory}
\subsection{BSE for Core-level Spectra}

Core-level x-ray absorption spectra (XAS), electron energy-loss spectra
(EELS),
and non-resonant inelastic x-ray scattering (NRIXS)  are all related
to the loss function
$L( \mathbf{q},\omega ) = - {\rm Im}\, {\epsilon^{-1}} ( \mathbf{q},\omega)$,
which is proportional to dynamic structure factor $S({\bf q},\omega)$.
Here ${\bf q}$ is the momentum transfer, while $\omega$ is the photon-energy
in XAS and the energy transfer in NRIXS and EELS.
(Unless otherwise specified we use Hartree atomic units
with $e=\hbar=m=1$ throughout this paper.)
For XAS we have $|{\bf q}| = \omega/c$, where $c\approx 137.036$ is the
speed of light.  Formally the loss function is given by
\begin{equation}
L( \mathbf{q},\omega ) 
 = - \frac{4 \pi }{ q^2}{\rm Im}\, \langle \Psi_0 \vert \hat{P}^\dagger 
[E_0 + \omega - \hat{H} + i\eta  ]^{-1}  \hat{P} \vert \Psi_0 \rangle.
 \label{epsinv}
\end{equation}
Here $\hat{P}$ is the operator that couples the many-body ground state
$|\Psi_0\rangle$ with the probe photon,
e.g., $e^{i \mathbf{q} \cdot \mathbf{r}}$ for NRIXS.  In the case of XAS, a slightly different formula provides the spectrum.  This formula involves a different operator, namely 
$ (\hat{\mathbf{e}} \cdot \mathbf{r}) + ( i /2 ) \, (\hat{\mathbf{e}} \cdot \mathbf{r}) \, (\mathbf{q} \cdot \mathbf{r}) + \cdots \,$, where $\hat{e}$ is the electric field direction.
$\hat H$ is the particle-hole Hamiltonian including self-energy and
lifetime effects as described below, and $\eta$ is a positive
infinitesimal.
Our approach for calculating Eq.\ (\ref{epsinv}) is adapted from the
treatments of the BSE in the quasi-particle approximation  of
Soininen and Shirley\cite{Soininen2001} and
Shirley,\cite{Shirley2005} which are summarized below.
Satellite effects due to multi-electron excitations are neglected,
though they can be approximated {\it a posteriori} in terms of a
spectral function in our self-energy approximation.

In order to evaluate the core-loss function in
Eq.\ (\ref{epsinv}), we make the approximation that the
excited states of a system 
can be described using a basis set of electron-hole states
$\{ \vert \Phi_{n \mathbf{k+q}, \alpha \mathbf{k}} \rangle \}$ .
Each electron-hole state
has an electron
in band $n$ with crystal momentum $\mathbf{k+q}$, and a
core hole denoted by an atomic level $\alpha$, with the
corresponding orbitals in each cell combined using a phased sum  
to form a Bloch state with crystal momentum $\mathbf{k}$.  
Our calculations also account for electron and hole spin degrees of
freedom, which are not explicitly written in what follows.  
The Hamiltonian $\hat H$ is represented by an effective particle-hole
Hamiltonian $\hat{H}_{\text{eff}}$
\begin{equation}
\hat{H}_{\text{eff}} = {\hat{H}}_e - {\hat{H}}_h + {\hat{H}}_{eh}.
\label{heff}
\end{equation}
Here ${\hat{H}}_{e/h}$ account for the single-particle energies
of the electron and hole states, including the spin-orbit interaction for the
core states and self-energy and/or lifetime corrections.
\begin{eqnarray}
\hat{H}_e &=& \hat H_0 + \Sigma, \nonumber \\
\hat{H}_h &=& \epsilon_c + i\Gamma,
\label{heh}
\end{eqnarray}
where $\hat{H}_0$ is the non-interacting single-particle Hamiltonian, 
$\epsilon_c$ is the core binding energy, $\Sigma$ is the final state self-energy and $\Gamma$ the core-hole
lifetime.
The electron-hole interaction is given by
\begin{equation}
{\hat{H}}_{eh} = \hat{V}_D + \hat{V}_X,
\label{H_eh}
\end{equation}
where
$\hat{V}_D$ and $\hat{V}_X$ account for the direct 
and exchange electron-hole interactions respectively.
The highly localized nature of core states allows for division of the
screening of the direct term $\hat V_D$ into a short-range part
calculated using the RPA, and a long-range part treated with a model dielectric
function,\cite{Shirley2006986}  while the exchange term $\hat V_X$ is
unscreened.  In order to treat multiplet effects the shortest range 
electron-core hole interaction is expressed using a compact Hamiltonian
based on the core atomic orbitals, PAW functions, 
and the related Slater $F$ and $G$ integrals.\cite{Shirley2005}

Inserting a complete set of
electron-hole states $ \{ \vert \Phi \rangle \} $, Eq.\ (\ref{epsinv}) becomes
\begin{eqnarray}
\label{epsinv2}
&&{\rm Im}\, [ \epsilon^{-1}(\mathbf{q},\omega)] = 
-\frac{4 \pi }{q^2} \sum_{\Phi , \Phi'} {\rm Im}\,
\left[ \langle \Psi_0 \vert \hat{P}^\dagger \vert \Phi\rangle\right. \nonumber\\
&& \qquad \left. \times \langle \Phi \vert
[ \omega - \hat{H}_{\text{eff}}  + i \Gamma ]^{-1} \vert \Phi' \rangle \langle \Phi' \vert \hat{P} \vert \Psi_0 \rangle \right] .
\end{eqnarray}
To obtain the particle-hole states, $\{ \vert \Phi \rangle \} $,
the band states $\phi_{n, \mathbf{k}+\mathbf{q}}(\mathbf{r})$
are calculated for both occupied and unoccupied levels
using the ground-state Kohn-Sham Hamiltonian, while the core states
$\chi_{\alpha,{\mathbf{k}}}(\mathbf{r})$ are taken to be
Bloch states derived from the atomic core states for a given absorption edge 
calculated with an atomic Hartree-Fock code.
A transition
$ \vert \Psi_0 \rangle 
\rightarrow 
\vert \Phi_{n{\mathbf{k}}+{\mathbf{q}},\alpha {\mathbf{k}}} \rangle$,
has a matrix element of the form  
\begin{eqnarray}
\langle \Phi_{n,\mathbf{k+q},\alpha\mathbf{k} }
\vert \hat{P} \vert \Psi_0 \rangle
&=& N^{-1} \sum_\mathbf{R} e^{i \mathbf{k \cdot R}}
\langle \phi_{n,\mathbf{k+q} } \vert \hat{P}
\vert \chi_{\alpha, \mathbf{R}} \rangle \nonumber \\
&=&   \langle \phi_{n, \mathbf{k+q}} \vert \hat{P}
\vert \chi_{\alpha, \mathbf{R=0} } \rangle .
\label{transitionmatrix}
\end{eqnarray}
Here N is the number of unit cells in the crystal. Making use of an appropriate sum over identical core 
states located at every lattice site $\{\mathbf{R}\}$, the particle-hole
state is thus constructed to have good crystal momentum $\mathbf{ k}$.
The resolvent in Eq.\ (\ref{epsinv2}) is then calculated using a
Lanczos algorithm.\cite{Shirley2005}

\subsection{PAW Transition Matrix Elements}

The transition matrix between the ground state and one-electron
excited states in Eq.~(\ref{transitionmatrix})
is then calculated locally by expressing the conduction band states
in terms of atomic states centered about the specific core-hole located
at position ${\bf \tau}$ within the unit cell using the projector augmented
wave (PAW) expansion
\cite{Blochl94}
\begin{equation}
\phi_{n,\mathbf{k}+\mathbf{q}}( \mathbf{R} +
\mathbf{r} + \mathbf{\tau}) \approx 
e^{i {\mathbf{(k+q)}} \cdot {\mathbf{R}} }
\sum_{\nu l m} A^{n {\mathbf{k}+\mathbf{q}}}_{\nu l m } 
F^{ps}_{\nu l} (r) Y_{l m} (\hat{\mathbf{r}}) ,
\end{equation}
where \{$F^{ps}_{\nu l}$\} are the PAW basis functions with
 angular quantum numbers l and m. The index $\nu$ denotes the use of 
multiple PAW basis functions per angular momentum to adequately span the space.
In terms of these PAW basis functions, transition matrix elements are found easily using an atomic structure program.   
Further details are given elsewhere.\cite{Soininen2001}

\subsection{Many-pole Self-energy}

One of the key considerations in calculations of core excitation spectra is the
treatment of the final-state self-energy shifts and damping effects observed
in experimental spectra.  A number of methods for this purpose
now exist.\cite{0953-8984-15-17-312} Because these effects are small
near an excitation threshold they are often neglected.
However, they become increasingly important when the excitation energy
exceeds plasmon and interband energies of order 10 eV,
and thus become crucial at high excitation energies.
To treat these effects we have implemented the efficient
many-pole GW self-energy model (MPSE) of Kas et al.  \cite{kas:195116} 
This MPSE is a straightforward extension of the
Hedin and Lundqvist single plasmon-pole GW self-energy model, and is based on
the calculated dielectric response specific to a given material, e.g., 
as obtained using AI2NBSE.

\begin{figure}
\begin{center}
\includegraphics[scale=0.30,clip,trim = 50 50 0 0]{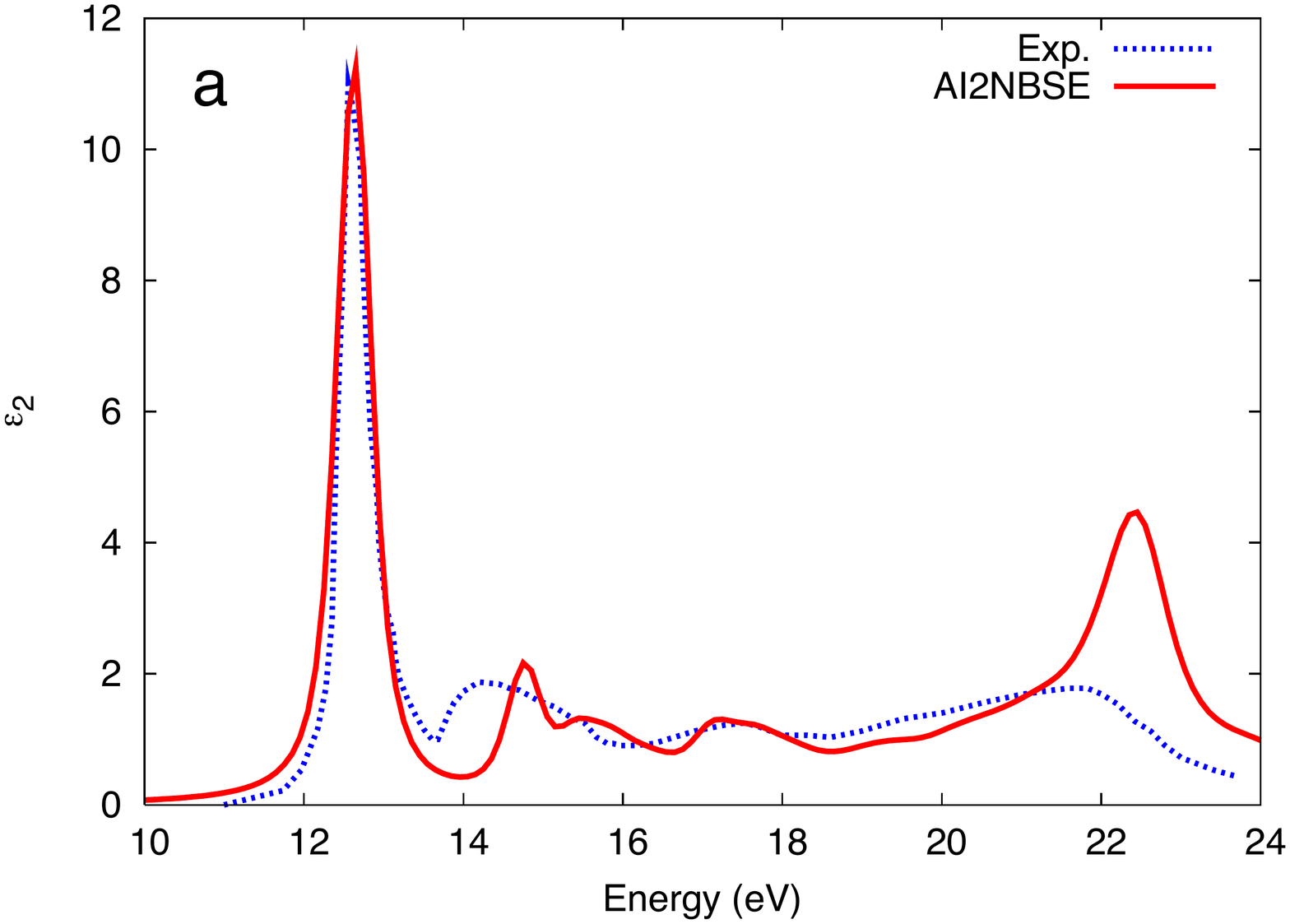}
\end{center}
\begin{center}
\includegraphics[scale=0.30,clip,trim = 52 50 0 0]{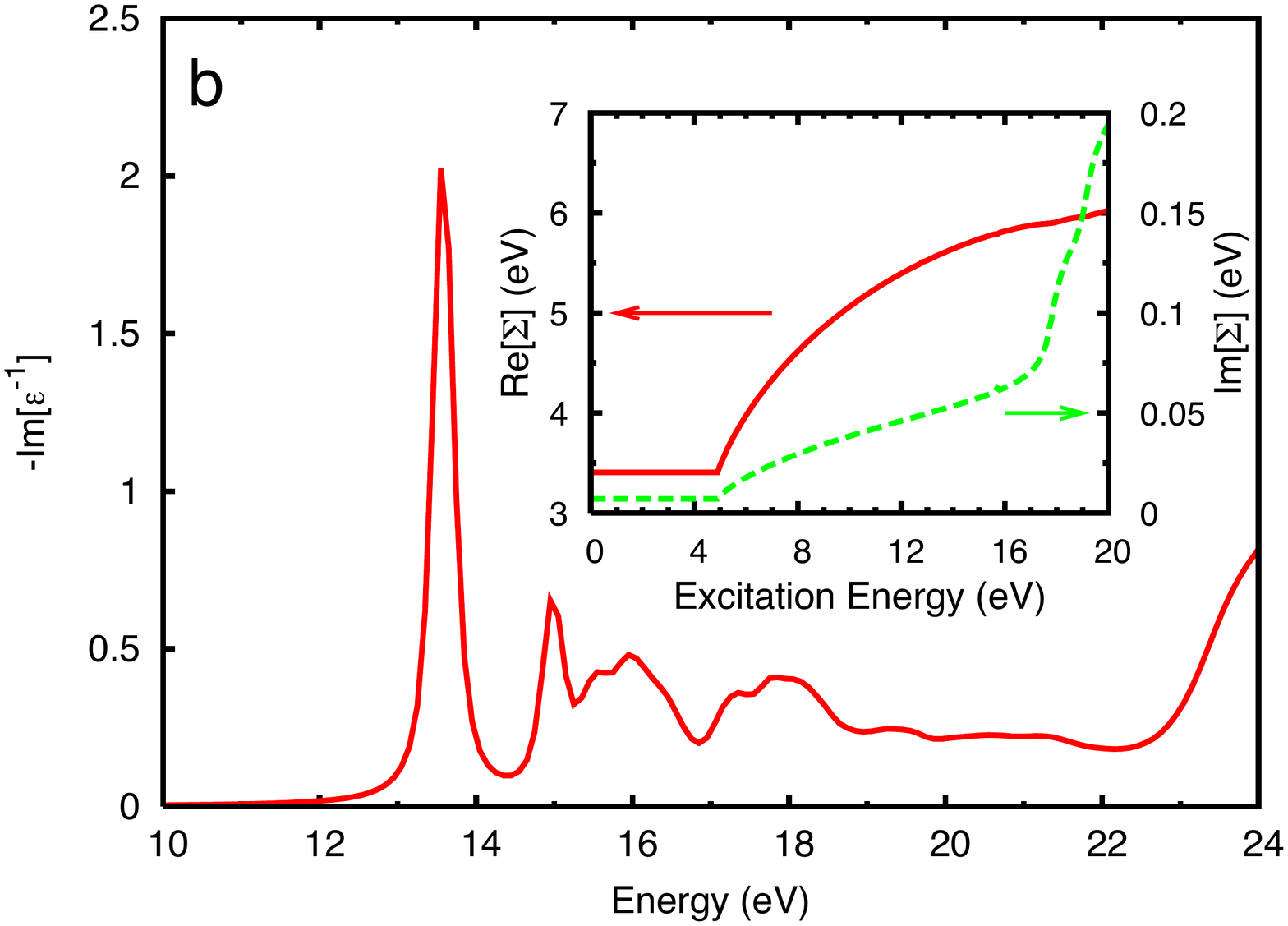}
\end{center}
\caption{ (Color online) Imaginary part of the dielectric
function (top) $\epsilon_2(\omega)$
calculated with AI2NBSE (red, solid line) and experiment\cite{Roessler1967}
(blue, dotted line) (data taken from Puschnig and Ambrosch-Draxl\cite{PhysRevB.66.165105});
and 
loss function (bottom) $-{\rm Im}\, \epsilon^{-1}(\omega)$ with (inset) real
and imaginary parts of the self-energy calculated from AI2NBSE. The
self-energy is shown relative to the valence band maximum, and the jump
in the real part at half the LDA gap is an artifact of the method.
This valence loss spectra is needed as input for the many-pole self energy
correction to the core spectra (see text).}
\label{lif_val}
\end{figure}

In brief, our implementation of the MPSE is as follows. We first represent
the loss function in the optical regime
$L(\omega) = -{\rm Im}[\epsilon^{-1}(q=0,\omega)]$
as a weighted sum over closely-spaced delta-functions at $\omega_j$
with weights $w_j = L(\omega_j)\Delta_j$ chosen to
preserve the loss-function
sum-rule.  The inclusion of many (typically of order 10$^2$) 
poles allows for smooth, energy
dependent broadening and quasi-particle shifts in excitation energies,
and accounts for losses in the near-edge region well below
the dominant excitations.
As an example, Fig.\ 1 shows a comparison of the imaginary part of
the dielectric function $\epsilon_2(\omega)$
for LiF as measured experimentally and that calculated by AI2NBSE. The pole
representation also permits a straightforward way to extrapolate
the long-wavelength limit $\epsilon^{-1}(q=0,\omega)$
to finite momentum transfer. Our MPSE uses the same the dispersion relation
as in the original Hedin-Lundqvist plasmon-pole model\cite{kas:195116}. 
We then calculate the self-energy $\Sigma=iGW$ within the GW approximation,
where the Green's function $G$ is taken to be the free propagator of an
electron gas and
$W = \epsilon^{-1}v$ is calculated using the many-pole model of
the dielectric function.
Using the same strategy as for the plasmon-pole model, the resulting
self-energy can be expressed as a Hartree-Fock exchange term, plus
a weighted sum of plasmon-pole like, dynamically screened exchange terms,
each with a different plasma frequency $\omega_j$. Thus the model retains the
efficiency of the plasmon-pole model yet yields accurate calculations
of the self-energy over a broad energy range from near the Fermi level to of order
10$^4$ eV above the Fermi level. A typical calculation can be performed in only a few minutes
on a modern single-processor desktop computer once the 
loss function at zero momentum transfer is known.
In order to apply the self-energy to calculations of core level XAS,
we use an \textit{a posteriori} convolution of the spectrum with a
Lorentzian spectral function built from our calculated quasi-particle
self-energy. The effects of multi-electron excitations 
can also be included in this convolution.\cite{kas:195116} 
 
Currently the MPSE is calculated only for the unoccupied
states. This self-energy is then added, assuming that all
transitions occur either from the highest occupied level in the case
of the valence response, or from a single localized core level in the
case of core response. For core excitations this is a good
approximation, while for valence response the approximation
is valid only for narrow valence bands and future refinements
should take the width of the valence band into account.
An example of the resulting XAS is shown for LiF in Fig.\  \ref{1}.
Clear improvement can be seen due to the MPSE, which corrects both
the peak positions and amplitudes in the spectrum.

\section{Calculations and Results}

In this section we illustrate our approach with a number of 
examples.  The results are sensitive to the choice of pseudopotentials
and PAW projectors, and care must be taken to ensure an adequate treatment.
Briefly, our implementation of OCEAN uses ABINIT wave-functions
with norm-conserving pseudopotentials generated using the
FHI\cite{Fuchs199967} or OPIUM\cite{opium} codes utilizing the designed
non-local approach; 
PAW transition elements from atomic core states utilizing from
4 to 10 projectors depending on valence band width;
the NIST NBSE solver for core excitations; and the
GW MPSE outlined above.  Our calculations use separate grids to calculate
the wave-functions for the final states and for the screening. For all of the
materials, calculations of the final states on a $10\times10\times10$
Brillouin-zone grid with a
symmetry breaking shift of (1/80, 2/80, 3/80) was found to be adequate for
convergence. The number of final state bands included in the calculation only
affects the range of validity of the spectrum above the edge. Depending on the
example, between 40 and 70 conduction bands are included. 

The screening in the direct interaction $\hat V_D$ was
calculated using states from a $2\times2\times2$ Brillouin-zone grid with a similar
symmetry breaking shift as the final state calculation. This
grid was found to be adequate for convergence. For the screening calculation,
a large number of unoccupied states must also be included for convergence.
Including bands up to 100~eV above the Fermi level was found to be
sufficient for all of our cases. Dielectric screening was calculated in
real-space as described by Shirley.\cite{Shirley2006986}

\subsection{LiF}

\begin{figure}
\begin{center}
\includegraphics[scale=0.30,clip,trim = 50 50 0 0]{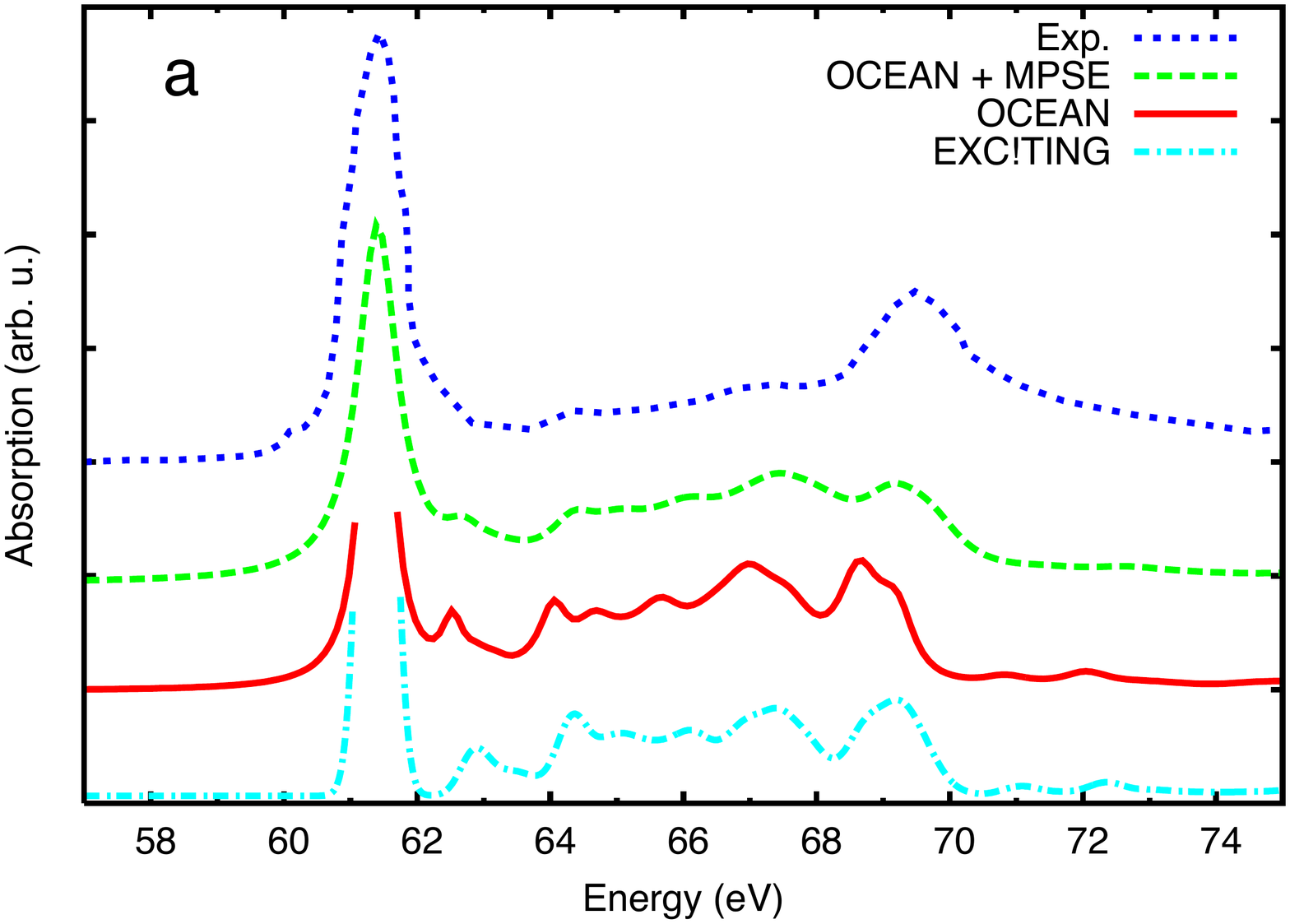}
\end{center}
\begin{center}
\includegraphics[scale=0.30,clip,trim = 50 50 0 0]{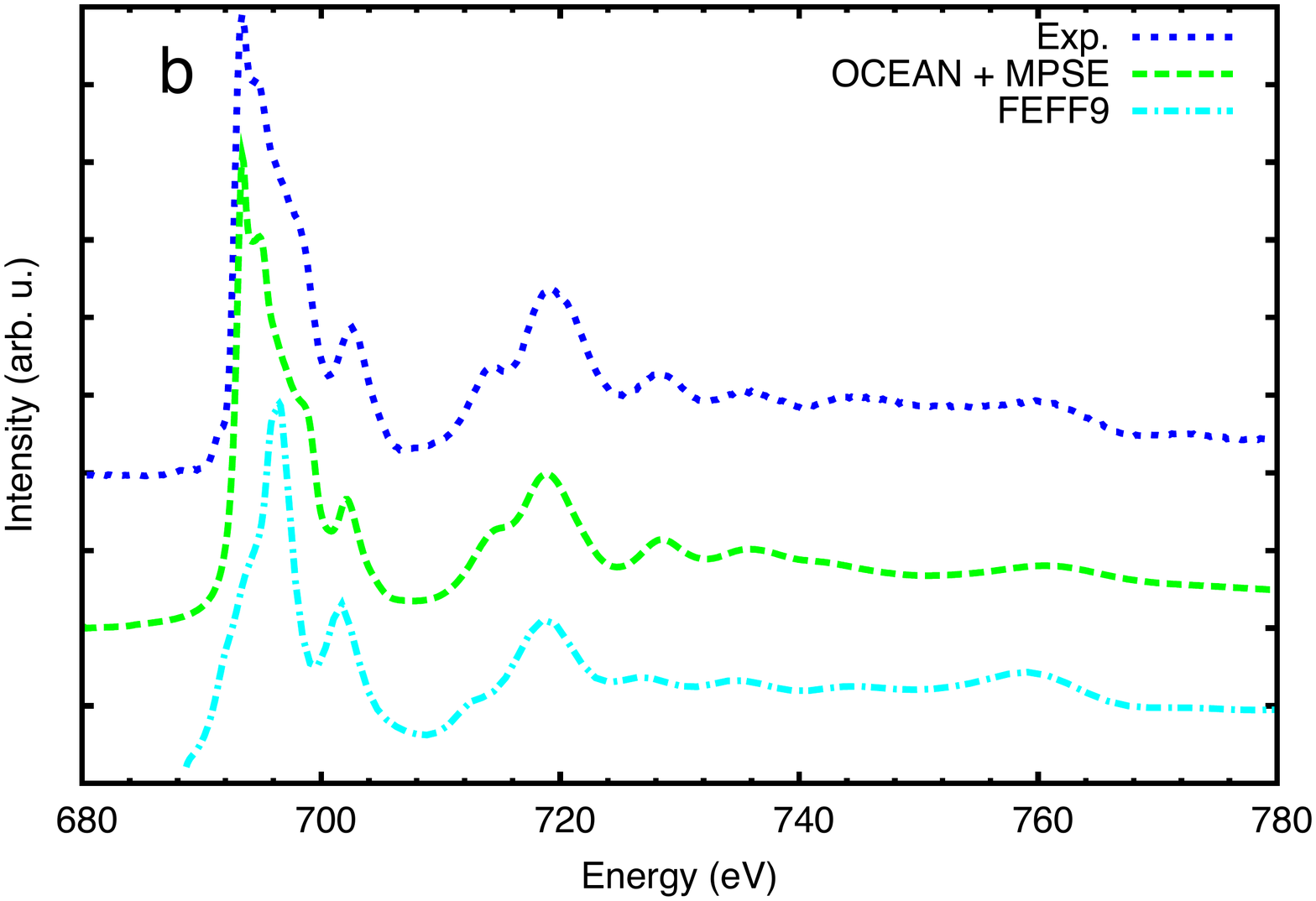}
\end{center}
\caption{ (Color online) a) The x-ray absorption near-edge structure (XANES) spectra for the Li K-edge of LiF (top)
and b) the F K-edge (bottom). The BSE spectra (solid, red line) are compared
to the same spectra convoluted with the MPSE correction
(green, dashed line) and experiment \cite{Handa2005,PhysRevB.49.3701} (Li data from
Olovsson et al.\cite{olovsson:041102}) (blue, dotted line).
In a) the result of another BSE code,\cite{olovsson:041102}
(light blue, dashed-dotted line) and in b) a calculation with the
FEFF9 code (light blue, dashed-dotted line) are shown for comparison.}
\label{1}
\end{figure}

LiF is a wide-gap insulator leading to strongly bound excitonic peaks in the
absorption spectra which are highly dependent on the screened direct
interaction between the excited electron and hole for their strength and
binding energy.  LiF has a rocksalt structure
with a lattice constant of 4.028~\AA,\cite{PhysRevB.52.8} 
and the electronic dielectric constant
$\epsilon_{\infty}$ (i.e., the value for low frequencies well above
characteristic vibrational frequencies) is 1.92.\cite{PhysRevB.58.9579} 
The first 56 conduction bands were included covering energies up to
100~eV above the Fermi level, and a 100~Ry cutoff was used for
the plane-wave basis.
While LDA calculations often underestimate band-gaps in insulators,
the MPSE almost completely corrects the conduction energy levels and
bandgap in LiF. This is illustrated in the UV calculation of
$\epsilon_2$ and the loss function
$-{\rm Im}\, \epsilon^{-1}$ (Fig.\ \ref{lif_val}).

Results for the Li K-edge XAS of LiF are presented in Fig.\ \ref{1}. 
Note that the inclusion of final-state broadening and self-energy shifts
from our MPSE yields results that agree well with 
experiment. If these final-state effects are ignored, one 
obtains sharper, more compressed spectra, in poorer agreement
with experiment.  However, our calculation
fails to account for the strength of the experimental peak near 70~eV,
i.e., about 10~eV above the edge. The reasons for this discrepancy are
not yet understood.  A comparison is also included with the results from
the BSE-based EXC!TING code.\cite{olovsson:041102} Although EXC!TING 
ignores energy dependent GW corrections, we find good agreement
in peak positions and spacing between the two codes for this case. 

Results for the F K-edge of LiF including the
final-state corrections from MPSE are shown in
Fig.\ \ref{1}b. As
for the Li K-edge, the broadening and self-energy shifts from the GW
MPSE lead to good agreement with experiment.
Overall the F K-edge (Fig.\ \ref{1}b) exhibits much better agreement with
the experiment than for the Li K-edge. A comparison is also included
with the FEFF9 code.\cite{rehrpccp} While the results are generally in good agreement,
our calculation yields better agreement with experiment in the edge
region.

\begin{figure}
\includegraphics[scale=0.30,clip,trim = 50 40 0 0]{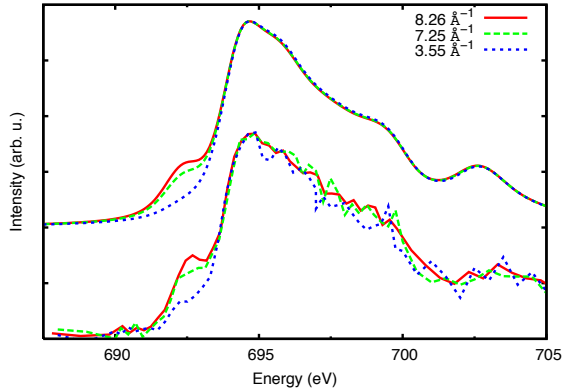}
\caption{ (Color online) The theoretical (above) and experimental\cite{PhysRevB.65.155111} (below) NRIXS spectra for LiF
for momentum transfer $q=3.55$~\AA$^{-1}$, 7.25~\AA$^{-1}$ and 8.26~\AA$^{-1}$ normalized to the 
height of the main peak. Note that
momentum transfer only affects the edge-spectra due to the behavior
of the $s$-type exciton in this system.}
\label{LiFnrixs}
\end{figure}

As an illustration of momentum transfer dependent calculations with
OCEAN, results for the NRIXS of LiF are shown in Fig.\  \ref{LiFnrixs}
for $q$ =3.55~\AA$^{-1}$, 7.25~\AA$^{-1}$ and 8.26~\AA$^{-1}$.
For this predominantly $s$-$p$ electron system the momentum transfer dependence
is strong only in the edge region, and reflects the behavior of an
$s$-type exciton. Our results are found to be in
good agreement both with experiment and with
previous calculations.

\subsection{KCl}

\begin{figure}
\begin{center}
\includegraphics[scale=0.30,clip,trim = 50 50 0 0]{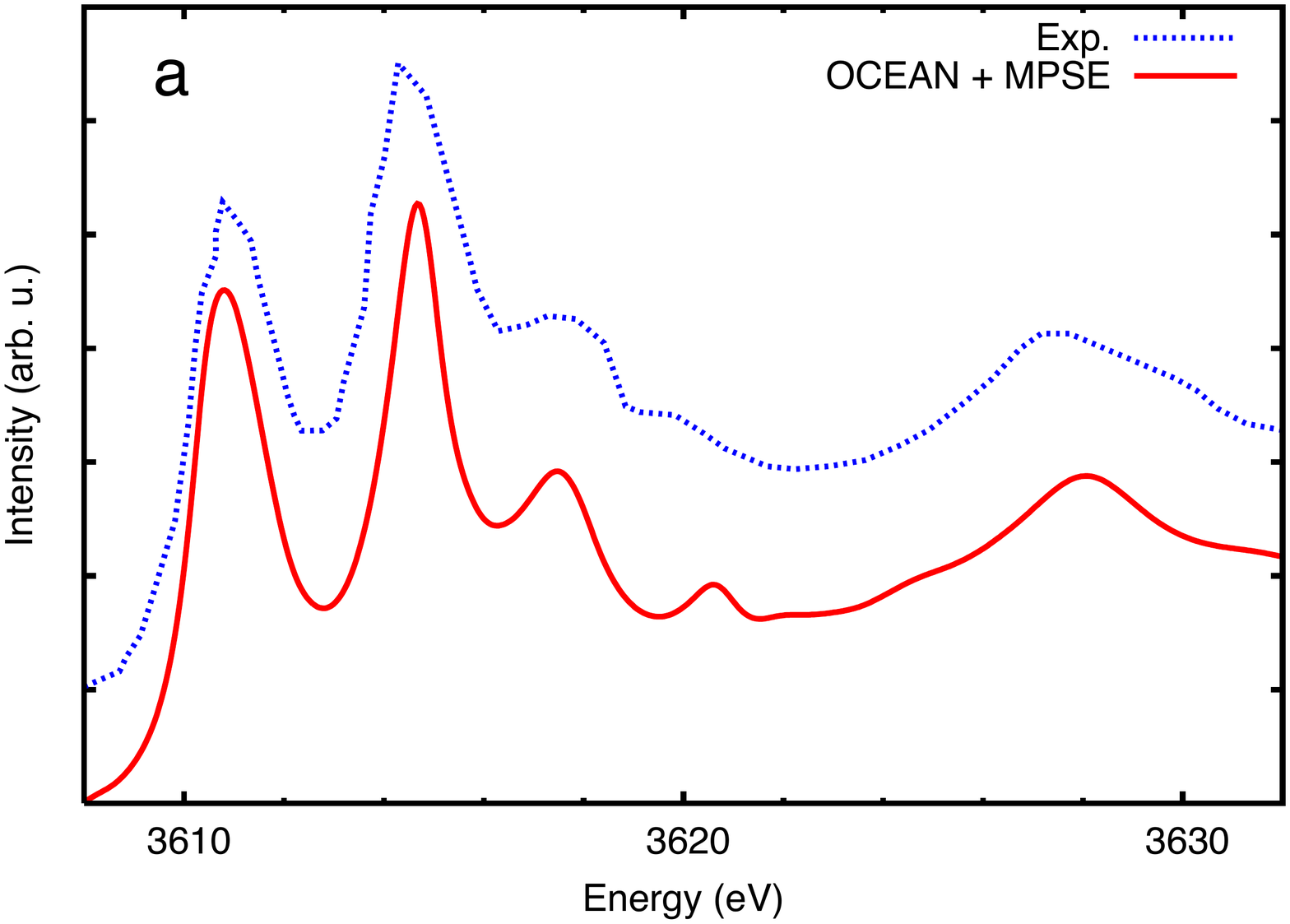}
\end{center}
\begin{center}
\includegraphics[scale=0.30,clip,trim = 50 50 0 0]{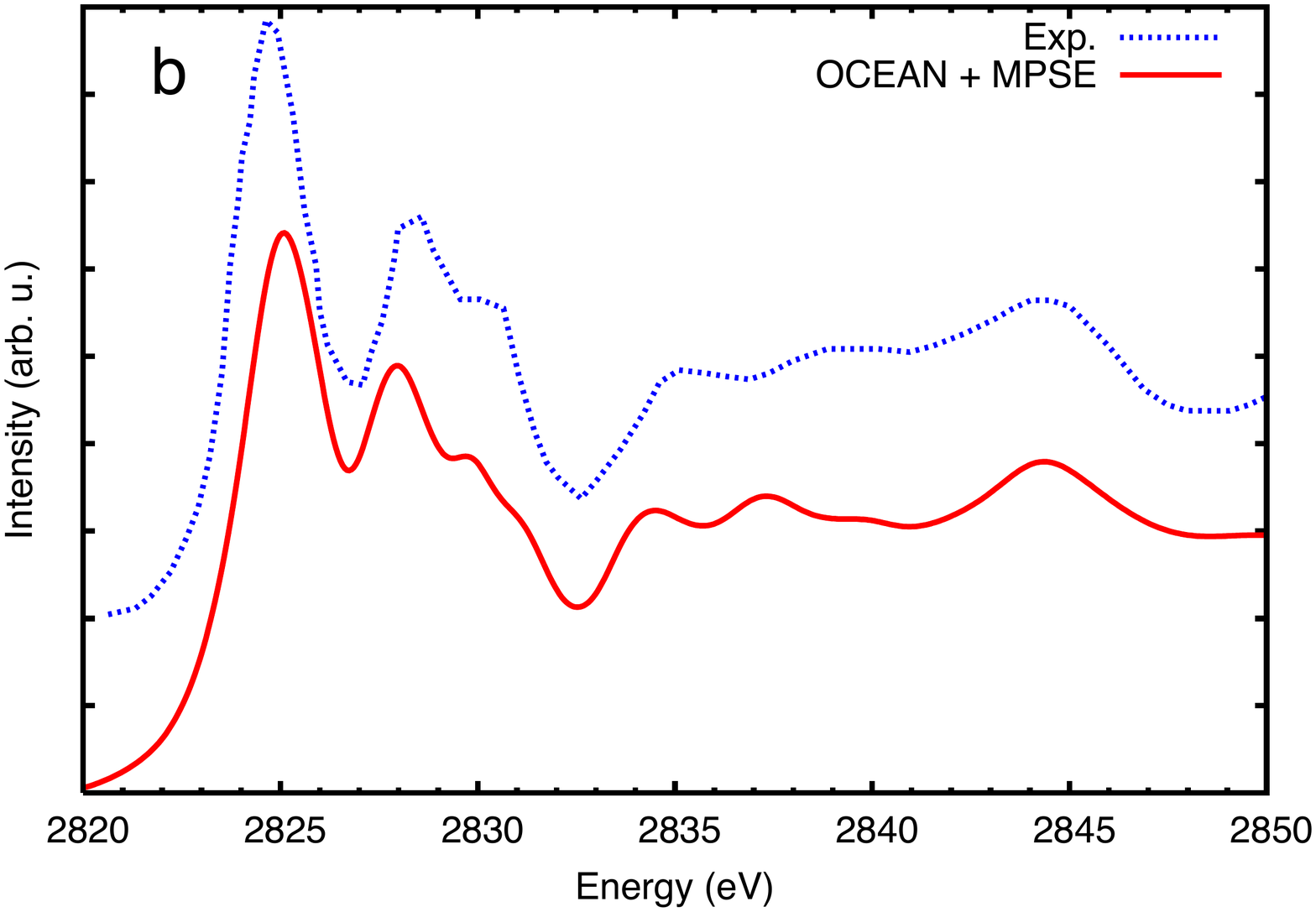}
\end{center}
\caption{ (Color online) a) The K-edge XAS of potassium (top) and b)
the Cl K-edge in KCl calculated with
OCEAN (red, solid line) and compared to experimental results
\cite{Lavrentyev1999} (blue, dotted line). }
\label{kcl_kk}
\end{figure}

As an example of deep core K-edge spectra we present the XAS for both
the K and Cl K-edges in KCl.
KCl has a rocksalt structure with lattice constant 6.29~\AA\, 
and $\epsilon_{\infty}$ = 2.19.\cite{PhysRevB.58.9579} A plane-wave
cut-off of 160~Ry\ and 172 conduction bands were used for the screening
calculations.  Both K-edges exhibit very good agreement
with experiment both for peak positions and intensities
(Fig. \ref{kcl_kk}).

\subsection{MgO}

\begin{figure}
\includegraphics[scale=0.30,clip,trim = 50 50 0 0]{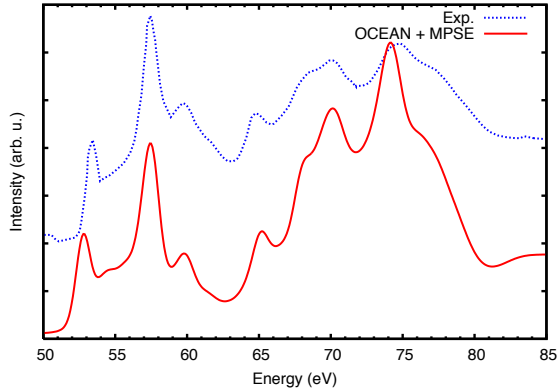}
\caption{ (Color online) Calculated XANES spectra for the Mg L$_{2,3}$-edge of
MgO compared with experimental reflection ($R^{{1}/{2}}$)
data.\cite{PhysRevB.44.1013}  The calculated spectra has been broadened and
by a Gaussian of 0.9 eV FWHM in addition to the MPSE.}
\label{mgoL23}
\end{figure}

To illustrate a shallow L-edge calculation we show the XAS for
the Mg L$_{2,3}$-edge in MgO (Fig.\ \ref{mgoL23}).
MgO also has a rocksalt structure with a lattice constant of 4.212~\AA\, and 
$\epsilon_{\infty}$ = 2.95.\cite{PhysRevB.58.9579} The screening
calculation used 196 conduction bands and a plane-wave cut-off of 200~Ry.
The splitting between the L$_2$ and L$_3$-edges 
in Mg is only 0.25~eV and is therefore hidden by
conduction band widths and experimental broadening.
Our XAS calculation of the Mg L$_{2,3}$-edge in MgO
is particularly sensitive to computational details such
as the treatment of core-hole screening. The calculation is
in good agreement for peak positions, though not the relative strengths.
The calculated spectrum is too weak in the range
60~eV to 70~eV while the peak near 75~eV is too strong and more narrow
than that in the experiment.  The excitonic peak is also too strong compared
to the main edge (Fig.\ \ref{mgoL23}).  It seems likely that under-screening
of the core hole leads to the overestimation of the strength of the
excitonic peak.

\subsection{SrTiO$_3$}

\begin{figure}
\includegraphics[scale=0.30,clip,trim = 50 50 0 0]{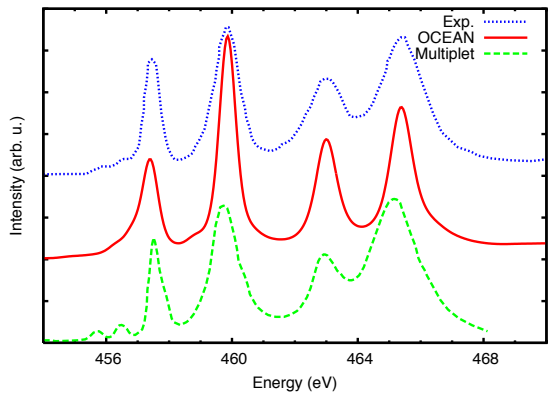}
\caption{(Color online) Calculated XANES spectra for the Ti L$_{2,3}$-edge
of SrTiO$_3$ (red, solid line) compared to experimental data \cite{Woicik07}
(blue, dotted line) and a multiplet calculation \cite{deGroot1994529}
(green, dashed line). Due to the close spacing of the edges the full
MPSE was not included.  The imaginary part of the self-energy was applied
by aligning the onset with respect to each edge and including a core-hole
lifetime (0.10~eV and 0.24~eV), shifting between the two at $461.5$ eV.  
This lifetime broadening neglects solid-state Coster-Kronig effects. }
\label{ti_l23}
\end{figure} 

The Ti L$_{2,3}$-edge of SrTiO$_3$ is included here as an example of multiplet effects.
Transition metal L$_{2,3}$-edges exhibit strong multiplet effects resulting 
in a shift in the  L$_3$/L$_2$ intensity ratio away
from the non-interacting ratio of 2:1. By including the core-level spin-orbit energy splitting
and simultaneously treating 2p$_{1/2}$ and 2p$_{3/2}$ initial states in the
BSE Hamiltonian, our approach yields {\it ab initio} estimates of shifts in the spectral weight.
Cubic perovskite SrTiO$_3$ has a lattice constant of 3.91~\AA\,\cite{PhysRevB.75.140103} and
$\epsilon_{\infty}$ = 5.82.\cite{dore1996} A plane-wave cut-off of 220~Ry and 450 conduction bands
were used for the screening.
Each Ti atom is surrounded by oxygen octahedra leading to a 2.5~eV splitting
in the 3d-like Ti final states according to their symmetry. Additionally
there is an approximately a 5.4~eV spin-orbit splitting between the L$_2$-
and L$_3$-edges.   Also the imaginary parts of the self-energy differed
for the states above each edge to account for the difference in core-hole
lifetimes between the L$_2$- and L$_3$-edges.
Our calculation gives good agreement for peak positions and produces a
ratio between the two edges in reasonable agreement with experiment
(Fig.\ \ref{ti_l23}). The two pre-edge features below 457~eV in the experiment
are reproduced in the calculation, though  they are too close to the
leading edge.

\section{Summary and Future Prospects}

We have developed a hybrid approach for GW/BSE calculations of core-excitation
spectra in periodic materials based on plane-wave pseudopotential wave
functions (from ABINIT in this case), PAW constructed matrix elements, the NIST BSE solver,
and a many-pole GW self energy from AI2NBSE. The method 
takes into account the many-body effects of inelastic losses and core-hole
interactions, and also includes atomic multiplet effects. 
This approach is implemented in the OCEAN package, which is applicable to
core level XAS, EELS and related spectra including finite momentum transfer.
However, the approach can be applied to many  other spectra, and some
future extensions include resonant inelastic x-ray scattering
(RIXS),\cite{Soininen2001} relativistic
EELS,\cite{Jorissen2009} and the mixed dynamic form factor (MDFF). 
The pseudopotential base
allows for computationally efficient calculations of fairly large systems
(currently up to about 50 atoms)
for near-edge spectra within about $10^2$~eV of threshold for arbitrary
core levels.  Moreover, by combining OCEAN with AI2NBSE and the complementary
RSGF code FEFF9, full-spectrum calculations from the UV-VIS to hard x-rays
are feasible.
Overall the results are found to be in good to excellent agreement with
experiment, and can improve on other theoretical approaches. 
Additional applications, e.g., to the XAS and NRIXS of water and ice,
are described elsewhere.\cite{ocean-ice}

\acknowledgments
This work was supported by DOE BES Grant DE-FG03-97ER45623 
and was facilitated by the DOE Computational Materials Science Network.
We thank C.\ Ambrosch-Draxl, X.\ Gonze, K. Hamalainen,
L.\ Reining, F.\ Vila, and the \textsc{ABINIT} development group
for comments and discussions.

\bibliographystyle{apsrev}
\bibliography{OCEAN4}

\end{document}